\documentclass[aps,pre,floats,preprint,showpacs,superscriptaddress]{revtex4}

\usepackage{epsfig}
\usepackage{latexsym,amsmath}

\begin{document}
\title{Interfaces and the edge percolation map of random directed networks}

\author{M. \'Angeles Serrano}

\author{Paolo De Los Rios}

\affiliation{Institute of Theoretical Physics, LBS, SB, EPFL, 1015
Lausanne, Switzerland}

\date{\today}

\begin{abstract}
The traditional node percolation map of directed networks is
reanalyzed in terms of edges. In the percolated phase, edges can
mainly organize into five distinct giant connected components,
interfaces bridging the communication of nodes in the strongly
connected component and those in the in- and out-components. Formal
equations for the relative sizes in number of edges of these giant
structures are derived for arbitrary joint degree distributions in
the presence of local and two-point correlations. The uncorrelated
null model is fully solved analytically and compared against
simulations, finding an excellent agreement between the theoretical
predictions and the edge percolation map of synthetically generated
networks with exponential or scale-free in-degree distribution and
exponential out-degree distribution. Interfaces, and their internal
organization giving place from ``hairy ball'' percolation landscapes
to bottleneck straits, could bring new light to the discussion of
how structure is interwoven with functionality, in particular in
flow networks.
\end{abstract}


\maketitle

\section{Introduction}
\label{Sec:intro}

The theory of percolation applied to random
networks~\cite{Dorogovtsev:2007a} has proven to be one of the most
notorious advances in complex networks
science~\cite{Albert:2002,Dorogovtsev:2003,Newman:2003}. Its
importance goes beyond the production in the short term of
theoretical results, which are general and relevant to systems in
many different fields. The implication are far-reaching. On one
hand, a number of different problems have a direct interpretation in
terms of percolation or can be mapped to it, such as the study of
resilience or vulnerability in front of random
failures~\cite{Havlin:2000} or SIR epidemic spreading
models~\cite{Grassberger:1983,Sander:2002,Sander:2003,Newman:2002b,Kenah:2007,Miller:2007}.
On the other hand, the emergent percolation landscape can strongly
affect properties such as fluency or navigability in self-organized
systems. Hence, the conformation of connectivity structures in the
percolated phase should ensure efficient communication at the global
level so that different parts of the system--individuals, modules,
or substructures- are able to interact for the whole to organize and
develop functionality.

In the case of undirected networks, where elements are linked by
channels operating in both directions, the basic percolation
discussion was centered around the appearance of a macroscopic
portion of connected nodes that are linked through undirected paths
and so can communicate among them. The critical point for the
appearance of this giant component and its relative size in number
of nodes and edges was
determined~\cite{Molloy:1995,Molloy:1998,Havlin:2000,Newman:2001b,Callaway:2000},
also in the presence of specific structural
attributes~\cite{Newman:2002a,Newman:2003a,Vazquez:2003,Dorogovtsev:2001,Krapivsky:2004,Schwartz:2002}.
In its turn, the standard picture in directed
graphs~\cite{Newman:2001b,Callaway:2000,Dorogovtsev:2001,Dorogovtsev:2001b,Boguna:2005,Serrano:2006a,Serrano:2006c}
establishes that this giant connected component may become much more
complex and internally organized in three main giant structures, the
in-component, the out-component, and the strongly connected
component, as well as other secondary aggregates such as tubes or
tendrils. This conformation, sometimes represented as a bow-tie
diagram~\cite{Broder:2000}, denotes a potential global flow -of
matter, energy, information...- organized around a core which
usually processes input into output.

In this work, we will see that the percolation landscape, the
aggregate of macroscopic connectivity structures in the percolated
phase above the critical point, is further shaped when edges, the
0-level primary building blocks of networks along with nodes, are
taken as starring elements. Five distinct components are found to be
relevant in the edge percolation map of directed networks, the
traditional strongly connected and the in and out node components,
and two newly identified interfaces bridging the communication
between them. In Sec.~\ref{Sec:comp}, we define the relevant
components and present analytical computations based on the
generating function formalism and the usual locally treelike
assumption for their relative size in number of edges in purely
directed random networks that can present local and two-point
correlations. In Sec.~\ref{Sec:uncorr}, the formal equations for the
most general situation will be reformulated for the prototypical
null model of uncorrelated networks. The corresponding analytical
results will be compared to simulations for networks with
exponential in and out degree distributions and to numerical
solutions associated to networks with scale-free in-degree and
exponential out-degree distributions. A discussion of the
implications coming out of this description will be provided in
Sec.~\ref{Sec:int}, where the concept of interface will be further
examined along indications of the potential relevance of its
internal structure, that could organize to produce from ``hairy
ball'' percolation landscapes to bottleneck straits. We end by
summarizing and giving some final remarks in Sec.~\ref{Sec:conc}.

\section{Edge components in directed networks}
\label{Sec:comp}

In the traditional node percolation map of directed networks the
core structure is the giant strongly connected component (GSCC),
where all vertices within can reach each other by a directed path.
When present, it serves as a connector of the giant in-component
(GIN), composed by all vertices that can reach the GSCC but cannot
be reached from it following directed paths, to the giant
out-component (GOUT), made of all vertices that are reachable from
the GSCC but cannot reach it following directed paths.

From the point of view of edges, the GIN and the GOUT unfold into
two structures each, the edge in-component (ICE) and the in
interface (ITF), and the edge out-component (OCE) and the out
interface (OTF) respectively, so that five giant components should
indeed be distinguished. This increase in the number of relevant
structures is a consequence of the fact that nodes are point objects
and they belong to just one of the three node components, whereas
edges can be considered as extended objects in the sense that they
could belong simultaneously to two different node components, having
for instance one end in the GIN or GOUT and the other in the GSCC.
This fact points to the necessity of defining new classes for edges.
We will not take into account aggregates such as tendrils or tubes,
so that edges will be classified into five different categories
depending on the affiliation of the nodes they are joining. Let us
recall that, in the node percolation map, the out- and in-components
of individual vertices are defined as the number of vertices (plus
itself), $s_i$, that are reachable from a given vertex and the
number of vertices (plus itself), $s_o$, that can reach that vertex,
respectively. The GSCC can be thus thought of as the set of vertices
with infinite in- and out-components simultaneously, and the GOUT
and GIN as the set of vertices with infinite in-component and
infinite out-component respectively, excluding the GSCC. Taking this
into consideration, we give the following definitions for the
different principal components of the edge percolation map of random
directed networks:

\begin{itemize}
\item The edge in-component, ICE, is the set of edges joining source and destination nodes
with finite in-component and infinite out-component. These edges are
connecting nodes within the GIN.
\item The in-interface, ITF, is the set of edges joining source nodes with finite in-component
and infinite out-component and destination nodes with infinite in-
and out-components. These edges are bridging the ICE and the SCE
(see below) by connecting nodes in the GIN to nodes in the SCC.
\item The edge strongly connected component, SCE, is the set of edges joining source and destination nodes with infinite
in- and out-components. These edges are connecting nodes within the
SCC.
\item The out-interface, OTF, is the set of edges joining source nodes with infinite in- and
out-components and destination nodes with infinite in-component and
finite out-component. These edges are bridging the SCE and the OCE
by connecting nodes in the SCC to nodes in the GOUT.
\item The edge out-component, OCE, is the set of edges joining source and destination nodes
with infinite in-component and finite out-component. These edges are
connecting nodes within the GOUT.
\end{itemize}
\begin{figure}[t]
\epsfig{file=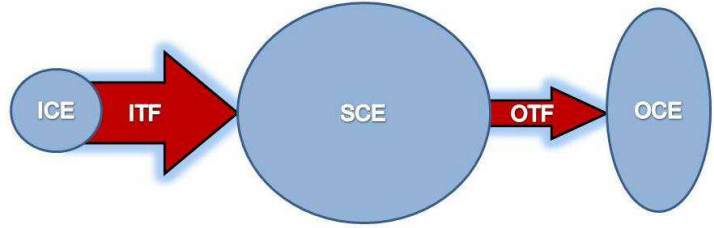, width=10cm} \caption{(color online).
Schematic representation of the main giant components in the edge
percolation map. As illustrated in the sketch, the different
components can be heterogeneous in their sizes.} \label{fig:S1}
\end{figure}

The critical point for the simultaneous appearance of the three
giant node components -as well as other secondary structures such as
tubes or tendrils- trivially marks also the emergence of the five
giant edge components. In the most general case, the condition
$\lambda_m>1$ characterizes the percolated phase, where $\lambda_m$
stands for the maximum eigenvalue of a characteristic matrix. In the
case of purely directed random networks, where the main attribute of
each node is its degree ${\bf k}\equiv(k_{i},k_{o})$ determined by
its incoming and outgoing number of connections $k_{i}$ and $k_{o}$,
the characteristic matrix in the presence of two-point correlations
was found to be $C_{{\bf k}{\bf k'}}^o$ (or $C_{{\bf k}{\bf k'}}^i$
with the same results)~\cite{Boguna:2005},
\begin{equation}
\begin{array}{rrr}
C_{{\bf k}{\bf k'}}^o&=&k'_o
P_o({\bf k'}|{\bf k})\\[0.3cm]
C_{{\bf k}{\bf k'}}^i&=&k'_i P_i({\bf k'}|{\bf k}),
\end{array}
\end{equation}
where the transition probabilities $P_{i}({\bf k'} | {\bf k})$ and
$P_{o}({\bf k'} | {\bf k})$ measure the likelihood to reach a vertex
of degree ${\bf k'}$ leaving from a vertex of degree ${\bf k}$ using
an incoming and an outgoing edge, respectively. If the degrees of
connected vertices are statistically uncorrelated, this condition
reduces to the first-born~\cite{Newman:2001b}
\begin{equation}
\sum_{k_{i},k_{o}}k_{o}(k_{i}-1)P(k_{i},k_{o}) > 0,
\label{criticalpoint}
\end{equation}
where $P(k_{i},k_{o})\equiv P({\bf k})$ is the joint degree
distribution of in- and out-degrees, that could encode local
correlations.

\subsection{Analytical computation of edge components size in purely directed networks}
In order to compute the sizes of the different giant components in
number of edges, the already traditional approach used in previous
developments is also appropriate with necessary adjustments. The
mathematical methodology is based on the generating function
formalism while the physical methodology explores the network with
branching processes which expand under the locally treelike
assumption~\cite{Newman:2001b,Dorogovtsev:2001b,Boguna:2005}.
Maximally random purely directed networks with local and two-point
correlations will be considered. This implies that the relevant
information about the topology of the network is encoded in the
joint degree distribution $P({\bf k},{\bf k'})$, where ${\bf k}$ is
the degree of the source node and {\bf k'} the degree of the
destination node, or, equivalently, in the degree distribution
$P({\bf k})$ along with the transition probabilities $P_{i}({\bf k'}
| {\bf k})$ and $P_{o}({\bf k'} | {\bf k})$. These are related
through the following degree detailed balance
condition~\cite{Boguna:2002,Boguna:2005}
\begin{equation}
k_o P({\bf k}) P_{o}({\bf k'} | {\bf k})=k'_i P({\bf k'}) P_{i}({\bf
k} | {\bf k'}), \label{detailed_dir}
\end{equation}
which is fulfilled whenever any edge leaving a vertex points to
another or, in other words, whenever the network is closed and does
not present dangling edge ends. Although the condition is satisfied
for the whole graph, the three node components -GIN, GSCC, and GOUT-
do not fulfill the detailed balance condition separately. If one
restricts to consider the nodes within the boundaries of each
component along all their connections, dangling ends can be found.
The interfaces are just the sets of edges that prevent the node
components from fulfilling the detailed balance condition
separately.

Apart from the distributions above, the calculations also rely on
the edge joint distribution $G(s_{i},s_{o};s'_{i},s'_{o})$
associated to directed edges joining source and destination
vertices. It measures the simultaneous occurrence of finite sizes
for the different single node components associated to the connected
vertices. More specifically, it measures the number of vertices
(plus itself), $s_{o}$, that are reachable from the source vertex
and the number of vertices (plus itself), $s_{i}$, that can reach
the source vertex, simultaneously to the number of vertices (plus
itself), $s'_{o}$, that are reachable from the destination vertex
and the number of vertices (plus itself), $s'_{i}$, that can reach
the destination vertex. Notice that if computations are done for
node components, the relevant distribution is $G(s_{i},s_{o})$ and
refers to just one node. According to the definitions above, and as
a function of the edge joint distribution, the relative sizes of the
different giant edge components can be formally written as
\begin{eqnarray}
g_{ice}&=&\sum_{s_{i}}\sum_{s'_{i}}G(s_{i},s_{o}=\infty;s'_{i},s'_{o}=\infty)\nonumber\\
g_{oce}&=&\sum_{s_{o}}\sum_{s'_{o}}G(s_{i}=\infty,s_{o};s'_{i}=\infty,s'_{o}) \nonumber\\
g_{itf}&=&\sum_{s_{i}}G(s_{i},s_{o}=\infty;s'_{i}=\infty,s'_{o}=\infty)\nonumber\\
g_{otf}&=&\sum_{s'_{o}}G(s_{i}=\infty,s_{o}=\infty;s'_{i}=\infty,s'_{o})\nonumber\\
g_{sce}&=&G(s_{i}=\infty,s_{o}=\infty;s'_{i}=\infty,s'_{o}=\infty),
\label{sizes0}
\end{eqnarray}
where we have made use of the fact that if the destination node has
an infinite out-component so it has the source node and,
analogously, if the in-component of the source node is infinite so
will be the in-component of the destination node. These functions
can be computed from the marginal distributions associated to
$G(s_{i},s_{o};s'_{i},s'_{o})$, which preserve just some of the four
variables. Their dependence on a given variable $s_{i/o}$ indicates
that the corresponding in or out-component of the source or
destination vertex (destination vertex with prima) is finite with
size $s_{i/o}$ regardless of the size of the rest of the involved
single node components. For instance, the function
$G(s_{i},;s'_{i},)$ measures the probability of an edge connecting a
source node with finite in-component of size $s_{i}$ to a
destination node with finite in-component of size $s'_{i}$,
regardless of the sizes of the out-components of connected nodes,
that could be finite or infinite (notice that for ease of notation
we just left blank the spaces corresponding to the marginalized
variables). In terms of these marginal probabilities, the relative
sizes of the main components are:
\begin{eqnarray}
g_{ice}&=&\sum_{s_{i},s'_{i}}G(s_{i},;s'_{i},)-\sum_{s_{i},s'_{i},s'_{o}}
G(s_{i},;s'_{i},s'_{o})\nonumber\\
g_{oce}&=&\sum_{s_{o},s'_{o}}G(,s_{o};,s'_{o})-\sum_{s_{i},s_{o},s'_{o}}
G(s_{i},s_{o};,s'_{o})\nonumber\\
g_{itf}&=&\sum_{s_{i}}G(s_{i},;,) - \sum_{s_{i},s'_{i}}
G(s_{i},;s'_{i},)-\nonumber\\
&& \sum_{s_{i},s'_{o}} G(s_{i},;,s'_{o})
+ \sum_{s_{i},s'_{i},s'_{o}}G(s_{i},;s'_{i},s'_{o})\nonumber\\
g_{otf}&=&\sum_{s'_{o}}G(,;,s'_{o}) - \sum_{s_{o},s'_{o}}
G(,s_{o};,s'_{o}) - \nonumber\\
&&\sum_{s_{i},s'_{out}} G(s_{i},;,s'_{o})
+ \sum_{s_{i},s_{o},s'_{o}}G(s_{i},s_{o};,s'_{o})\nonumber\\
g_{sce}&=&1-\sum_{s_{i}}G(s_{i},;,) - \sum_{s'_{o}} G(,;,s'_{o})
+\nonumber\\
&&\sum_{s_{i},s'_{o}} G(s_{i},;,s'_{o}). \label{sizes1}
\end{eqnarray}
These marginal probabilities depend on the degrees of the nodes at
the ends of the edge under consideration. Edges connecting nodes in
the same degree classes will be considered statistically equivalent,
so that these functions should be rewritten over joint degree
classes. For instance,
\begin{equation}
G(s_{i},;s_{i},s'_{o})=\sum_{{\bf k},{\bf k'}}P({\bf k},{\bf
k'})G(s_{i},;s_{i},s'_{o}|{\bf k},{\bf k'}),
\end{equation}
and analogously for the rest. To calculate these conditional
probabilities we have to introduce at this point the probability
functions $g_{o}(s|{\bf k})$ and $g_{i}(s|{\bf k})$, which represent
the distributions of the number of reachable vertices from a vertex,
given that we have arrived to it from another source vertex of
degree ${\bf k}$ following one of its outgoing or incoming edges,
respectively. These functions are exactly the same as those already
introduced in previous works for the computation of the sizes of the
GIN, GOUT and GSCC. The marginal conditional probabilities can then
be expressed as functions of these single-node probabilities, that
in its turn obey closed equations obtained from an iterative
procedure which applies the techniques of random branching processes
under the locally treelike assumption. This hypothesis is correct if
the length of cycles present in the network is of the order of its
diameter, so that the sizes of single node components can be exposed
by subsequent jumps from neighbors to neighbors of neighbors without
returning to already visited ones (the presence of lower order loops
would induce overcounting). In this way, the problem can be formally
solved in the general correlated case.

As a way of example, it will suffice here to provide the expression
of one of the marginal conditional distributions as a function of
$g_{o}(s|{\bf k})$ and $g_{i}(s|{\bf k})$ to illustrate the
derivation. Assuming the locally treelike condition, one of the two
relevant marginal conditional probabilities in the computation of
the ICE can be written as
%\begin{widetext}
\begin{eqnarray}
G(s_{i},;s'_{i},s'_{o}|{\bf k},{\bf k'})&=&\sum_{s_1^{i}\cdots
s_{k_i}^{i}} g_{i}(s_1^{i}|{\bf k}) \cdots g_{i}(s_{k_i}^{i}|{\bf
k}) \delta_{s_1^{i}+\cdots+s_{k_i}^{i}+1,s_{i}} \nonumber\\
&\times& \sum_{s_1^{'i}\cdots s_{k'_i-1}^{'i}} g_{i}(s_1^{'i}|{\bf
k'}) \cdots g_{i}(s_{k'_i-1}^{'i}|{\bf k'})
\delta_{s_i+s_1^{'i}+\cdots+s_{k_i-1}^{'i}+1,s'_{i}}  \nonumber\\
&\times& \sum_{s_1^{'o}\cdots s_{k'_o}^{'o}} g_{o}(s_1^{'o}|{\bf
k'}) \cdots g_{o}(s_{k'_o}^{'o}|{\bf k'})
\delta_{s_1^{'o}+\cdots+s_{k_o}^{'o}+1,s'_{o}}. \label{eq:gorda}
\end{eqnarray}
%\end{widetext}
This expression for the joint multi-component conditional size
distribution $G(s_{i};s'_{i},s'_{o}|{\bf k},{\bf k'})$ needs three
simultaneous computations: the number of vertices that can reach the
source node, the number of vertices that can reach the destination
node, and the number of nodes that the destination node can reach
itself. The procedure starts from an edge linking nodes of degrees
${\bf k}$ and ${\bf k'}$ and splits the sets $s_{i}$, $s'_{i}$ and
$s'_{o}$ into the different contributions associated to the
corresponding neighbors. For instance, the number of edges that
bring to the degree-${\bf k}$ source node, $s_i$, can be computed as
the sum of the different contributions that can reach each of its
$k_i$ incoming neighbors, $s^i_1+\cdots+s^i_{k_i}$. This corresponds
to the first set of summations of the three that appear in
Eq.~(\ref{eq:gorda}). Independent equations for the functions $g_i$
and $g_o$ can be found by expanding iteratively this procedure:
\begin{eqnarray}
g_{i}(s|{\bf k})&=&\sum_{\bf k'}P_i({\bf k'}|{\bf
k})g_{i}(s_{1}|{\bf k'})\cdots g_{i}(s_{k'_i}|{\bf k'})
\delta_{S_{k'_i},s}\nonumber\\
g_{o}(s|{\bf k})&=&\sum_{\bf k'}P_o({\bf k'}|{\bf
k})g_{o}(s_{1}|{\bf k'})\cdots g_{o}(s_{k'_o}|{\bf k'})
\delta_{S_{k'_o},s}, \label{gs0}
\end{eqnarray}
where $S_{k'_i}=s_1+\cdots+s_{k'_i}+1$ and
$S_{k'_o}=s_1+\cdots+s_{k'_o}+1$. These equations become tractable
using the generating function formalism. In mathematical terms,
generating functions are obtained by applying the transformation
$\hat{f}(z)\equiv \sum_s f(s) z^s$, so that functions are brought to
the discrete Laplace space. Once transformed for the variables $s$,
Eqs.~(\ref{gs0}) become closed for $\hat{g}_i$ and $\hat{g}_o$,
\begin{eqnarray}
\hat{g}_{i}(z|{\bf k})&=& z \sum_{\bf k'} P_i({\bf k'}|{\bf k})\hat{g}_{i}(z|{\bf k'})^{k'_i}\nonumber\\
\hat{g}_{o}(z|{\bf k})&=& z \sum_{\bf k'} P_o({\bf k'}|{\bf
k})\hat{g}_{o}(z|{\bf k'})^{k'_o}. \label{gs1}
\end{eqnarray}
All summations over finite sizes of the joint conditional size
distributions correspond to their generating functions evaluated at
$z=1$. Eventually, those depend on $\hat{g}_{i}(1|{\bf k})$ and
$\hat{g}_{i}(1|{\bf k})$:
\begin{eqnarray}
\hat{G}(1,;1,1)&=& \sum_{{\bf k},{\bf k'}} P({\bf k},{\bf k'})\hat{g}_{i}(1|{\bf k})^{k_i}\hat{g}_{i}(1|{\bf k'})^{k'_i-1}\hat{g}_{o}(1|{\bf k'})^{k'_o}\nonumber\\
\hat{G}(1,1;,1)&=& \sum_{{\bf k},{\bf k'}} P({\bf k},{\bf
k'})\hat{g}_{i}(1|{\bf k})^{k_i}\hat{g}_{o}(1|{\bf
k})^{k_o-1}\hat{g}_{o}(1|{\bf k'})^{k'_o}\nonumber\\
\hat{G}(,1;,1)&=& \sum_{{\bf k},{\bf k'}} P({\bf k},{\bf k'})\hat{g}_{o}(1|{\bf k})^{k_o-1}\hat{g}_{o}(1|{\bf k'})^{k'_o}\nonumber\\
\hat{G}(1,;1,)&=& \sum_{{\bf k},{\bf k'}} P({\bf k},{\bf k'})\hat{g}_{i}(1|{\bf k})^{k_i}\hat{g}_{i}(1|{\bf k'})^{k'_i-1}\nonumber\\
\hat{G}(1,;,1)&=& \sum_{{\bf k},{\bf k'}} P({\bf k},{\bf k'})\hat{g}_{i}(1|{\bf k})^{k_i}\hat{g}_{o}(1|{\bf k'})^{k'_o}\nonumber\\
\hat{G}(,;,1)&=& \sum_{{\bf k},{\bf k'}} P({\bf k},{\bf k'})\hat{g}_{o}(1|{\bf k'})^{k'_o}\nonumber\\
\hat{G}(1,;,)&=& \sum_{{\bf k},{\bf k'}} P({\bf k},{\bf
k'})\hat{g}_{i}(1|{\bf k})^{k_i}. \label{subgcorr}
\end{eqnarray}
These expressions will allow us to compute easily the relative sizes
of the different components:
\begin{eqnarray}
g_{ice}&=&\hat{G}(1,;1,)-\hat{G}(1,;1,1) \nonumber\\
g_{oce}&=&\hat{G}(,1;,1)-\hat{G}(1,1;,1) \nonumber\\
g_{itf}&=&\hat{G}(1,;,)-\hat{G}(1,;1,)-\hat{G}(1,;,1)+\hat{G}(1,;1,1)\nonumber\\
g_{otf}&=&\hat{G}(,;,1)-\hat{G}(,1;,1)-\hat{G}(1,;,1)+\hat{G}(1,1;,1)\nonumber\\
g_{sce}&=&1-\hat{G}(1,;,)-\hat{G}(,;,1)+\hat{G}(1,;,1).
\label{gcorr}
\end{eqnarray}
Notice that the sizes of the interfaces can also be written as
\begin{eqnarray}
g_{itf}&=&\hat{G}(1,;,)-\hat{G}(1,;,1)-g_{ice}\nonumber\\
g_{otf}&=&\hat{G}(,;,1)-\hat{G}(1,;,1)-g_{oce}.\nonumber\\
\end{eqnarray}

The set of Eqs.~(\ref{gs1})-(\ref{subgcorr})-(\ref{gcorr})
determines completely the relative sizes in number of edges of the
main giant components of the edge percolation map of two-point
correlated purely directed networks.

\section{Uncorrelated purely directed networks} \label{Sec:uncorr}

The formal solution given in the previous section becomes simpler
for the classical null model of uncorrelated networks. This will
allow us to perform further analytical computations that will be
checked against simulation results in order to contrast the accuracy
of the theory.

The absence of two-point correlations make possible to factorize the
joint degree distribution, and the conditional degree distributions
also simplify:
\begin{equation}
 P({\bf k},{\bf k'})=\frac{k'_i k_oP({\bf k})P({\bf k'})}{\langle k_i
 \rangle^2}
\label{transition_uncorrelated}
\end{equation}and
\begin{equation}
 P_{o}({\bf k'} | {\bf k})=\frac{k'_i P({\bf k'})}{\langle k_i \rangle} \mbox{ , }
 P_{i}({\bf k'} | {\bf k})=\frac{k'_o P({\bf k'})}{\langle k_i
 \rangle}.
\label{transitioninout_uncorrelated}
\end{equation}
In this situation, Eqs.~(\ref{gs1}) evaluated in $z=1$ reduce to
\begin{eqnarray}
\hat{g}_{i}(1|{\bf k})&\equiv&\hat{g}_{i}(1)= \sum_{\bf k} \frac{k_oP({\bf k})}{\langle k_i \rangle}\hat{g}_{i}(1)^{k_i}\nonumber\\
\hat{g}_{o}(1|{\bf k})&\equiv&\hat{g}_{o}(1)= \sum_{\bf k}
\frac{k_iP({\bf k})}{\langle k_i \rangle}\hat{g}_{o}(1)^{k_o},
\label{uncorrgs1}
\end{eqnarray}
so that the relative sizes in number of edges of the different
components in the uncorrelated case just depend on the joint degree
distribution $P({\bf k})$ and the single-node in and out generating
functions $\hat{g}_{i}(1)$ and $\hat{g}_{o}(1)$, and can be written
as
\begin{eqnarray}
g_{ice}&=&\sum_{\bf k}\frac{k_iP({\bf k})}{\langle k_i \rangle}\hat{g}_{i}(1)^{k_i}(1-\hat{g}_{o}(1)^{k_o})\nonumber\\
g_{oce}&=&\sum_{\bf k}\frac{k_oP({\bf k})}{\langle k_i \rangle}\hat{g}_{o}(1)^{k_o}(1-\hat{g}_{i}(1)^{k_i}) \nonumber\\
g_{itf}&=&\hat{g}_i(1)(1-\hat{g}_o(1))-g_{ice}\nonumber\\[0.2cm]
g_{otf}&=&\hat{g}_o(1)(1-\hat{g}_i(1))-g_{oce}\nonumber\\[0.2cm]
%g_{itf}&=&\hat{g}_i(1)(1-\hat{g}_o(1))-\nonumber\\
%&&\sum_{\bf k}\frac{k_iP({\bf k})}{\langle k_i \rangle}\hat{g}_{i}(1)^{k_i}(1-\hat{g}_{o}(1)^{k_o})\nonumber\\
%g_{otf}&=&\hat{g}_o(1)(1-\hat{g}_i(1))-\nonumber\\
%&&\sum_{\bf k}\frac{k_oP({\bf k})}{\langle k_i \rangle}\hat{g}_{o}(1)^{k_o}(1-\hat{g}_{i}(1)^{k_i})\nonumber\\
g_{sce}&=&(1-\hat{g}_i(1))(1-\hat{g}_o(1)). \label{uncorrsizes0}
\end{eqnarray}
If local correlations are also absent, the expressions above become
even simpler
\begin{eqnarray}
g_{ice}&=&(1-\hat{g}_{o}(1))\sum_{k_i}\frac{k_i\hat{g}_{i}(1)^{k_i}P(k_i)}{\langle k_i \rangle}\nonumber\\
g_{oce}&=&(1-\hat{g}_{i}(1))\sum_{k_o}\frac{k_o\hat{g}_{o}(1)^{k_o}P(k_o)}{\langle k_i \rangle} \nonumber\\
g_{itf}&=&(1-\hat{g}_{o}(1))\hat{g}_i(1)-g_{ice}\nonumber\\[0.2cm]
g_{otf}&=&(1-\hat{g}_{i}(1))\hat{g}_o(1)-g_{oce}\nonumber\\[0.2cm]
%g_{itf}&=&(1-\hat{g}_{o}(1))(\hat{g}_i(1)-\sum_{k_i}\frac{k_i\hat{g}_{i}(1)^{k_i}P(k_i)}{\langle k_i \rangle})\nonumber\\
%g_{otf}&=&(1-\hat{g}_{i}(1))(\hat{g}_o(1)-\sum_{k_o}\frac{k_o\hat{g}_{o}(1)^{k_o}P(k_o)}{\langle k_i \rangle})\nonumber\\
g_{sce}&=&(1-\hat{g}_i(1))(1-\hat{g}_o(1)) \label{sizes00}
\end{eqnarray}
with
\begin{eqnarray}
\hat{g}_{i}(1)= \sum_{k_i}P(k_i)\hat{g}_{i}(1)^{k_i}\nonumber\\
\hat{g}_{o}(1)= \sum_{k_o}P(k_o)\hat{g}_{o}(1)^{k_o}. \label{gs00}
\end{eqnarray}

\subsection{Comparing against simulations}
In order to ascertain the accuracy of the theory, we contrast the
analytical results with those obtained from simulating uncorrelated
purely directed networks with given joint degree distribution of the
form $P({\bf k})=P(k_i)P(k_o)$. Uncorrelated networks are generated
according to a slightly modified version of the Molloy-Reed
prescription~\cite{Molloy:1995,Molloy:1998} -which is based on the
configuration model~\cite{Bender:1978,Bollobas:1980} and constructs
maximally random networks with a given degree sequence- to produce
directed connections controlling that $\sum_i k_i=\sum_o k_o$ and
also taking care of avoiding multiple or self-connections.

\subsubsection{Exponential in- and out-degree distributions}

For the first case study, we chose $P(k_i)$ and $P(k_o)$ of the form
\begin{equation}
P(k)=\left\{
\begin{array}{lr}
P_0 & k=0\\[0.5cm]
\displaystyle{\frac{(1-P_0)^2}{\langle k \rangle}
}\left[1-\frac{(1-P_0)}{\langle k \rangle}\right]^{k-1} & k \ge 1
\end{array}
\right.,\label{expPk}
\end{equation}
so that a full analytical solution is available. The sizes of the
giant components in the edge percolation map for this particular
joint degree distribution just depend on the parameters $P_0^{i}$
and $P_0^{o}$ and the average degree $\langle k_i \rangle=\langle
k_o \rangle$. Substituting Eqs.~(\ref{expPk}) into
Eqs.~(\ref{gs00}), the solutions are found to be
\begin{equation}
\hat{g}_i(1)=\frac{P_0^{i}}{q^{i}}\mbox{ ,
}\hat{g}_o(1)=\frac{P_0^{o}}{q^{o}}\mbox{ ,
}q=1-\frac{1-P_0}{\langle k_i \rangle}, \label{sol:exp}
\end{equation}
and the relative sizes
\begin{eqnarray}
g_{ice}&=&\frac{\hat{g}_i(1)(1-\hat{g}_o(1))}{\langle k_i \rangle^2}\nonumber\\
g_{oce}&=&\frac{\hat{g}_o(1)(1-\hat{g}_i(1))}{\langle k_i \rangle^2}\nonumber\\
g_{itf}&=&\frac{\hat{g}_i(1)(1-\hat{g}_o(1))}{\langle k_i
\rangle^2}(\langle k_i \rangle^2-1)\nonumber\\
g_{otf}&=&\frac{\hat{g}_o(1)(1-\hat{g}_i(1))}{\langle k_i \rangle^2}(\langle k_i \rangle^2-1)\nonumber\\
g_{sce}&=&(1-\hat{g}_i(1))(1-\hat{g}_o(1)). \label{sizesexp}
\end{eqnarray}
\begin{figure}[t]
 \epsfig{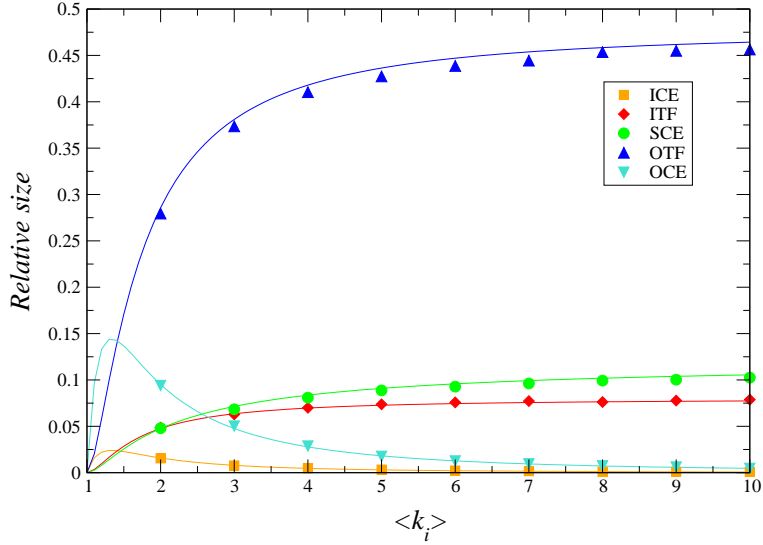}
  \caption{(color online). Relative sizes of the main giant components in the edge percolation map of networks with exponential
  in- and out-degree distributions as a function of the average degree. Simulation results (dots)
  correspond to 1-realization measures on synthetic networks with $N=10^5$
  vertices, $P_0^{i}=0.4$, and $P_0^{o}=0.8$. Solid lines are the analytical solutions Eqs.~(\ref{sol:exp})-(\ref{sizesexp}).}
  \label{fig:1}
\end{figure}
We compared these results with direct measures of the edge
components on a synthetic set of purely directed random networks
with $N=10^5$. We fix the values $P_0^{i}=0.4$, $P_0^{o}=0.8$, and
vary the average degree from $\langle k_i \rangle=1$ to $\langle k_i
\rangle=10$. As Fig.~\ref{fig:1} shows, the conformity of our
formulas to the simulation results is excellent. Notice also that
for this particular choice of the parameters $P_0^{i}$ and
$P_0^{o}$, the out-interface, OCE, is by far the biggest edge
component in the percolated phase for all values of the average
degree above approximately $1.5$ , followed with a noticeable
difference by the edge strongly connected component, SCE, and the in
interface, ICE. In this example, the interfaces are much stronger
than the edge in- and out-components, practically absent for high
degrees. This edge percolation map is seen to be quite stable for
most of the average degree range (see Sec.~{\ref{Sec:int}} for
further discussion).

\subsubsection{Scale-free in-degree and exponential out-degree
distributions}
\begin{figure}[t]
\epsfig{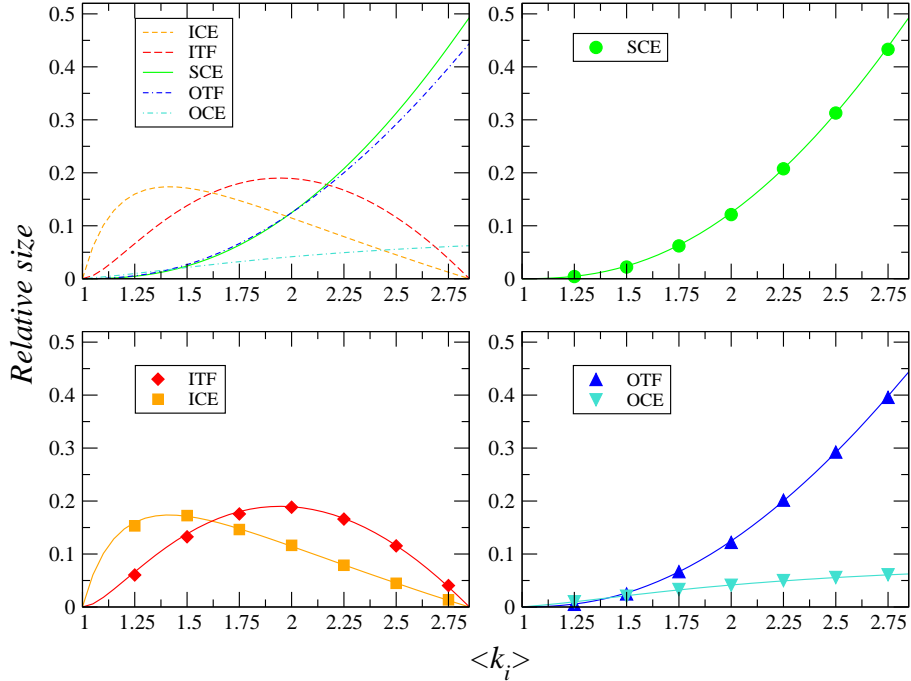}
  \caption{(color online). Relative sizes of the giant components in the edge percolation map of uncorrelated
  networks with scale-free in-degree distribution, $\gamma=2.2$, and exponential
  out-degree distribution, $P_0^{out}=0.4$, as a function of the average degree. Simulation results (dots)
  correspond to synthetic networks of size $N=5\cdot10^5$ vertices, 3 realizations for the first two point and one realization
  for the rest. Solid lines correspond to the numerical solutions of Eqs.~(\ref{sizes00})-(\ref{gs00}).}
  \label{fig:2}
\end{figure}

In some real networks, such as the WWW~\cite{Serrano:2007a} for
instance, the in-degree distribution exhibits a heavy-tailed form
well approximated by a power-law behavior $P(k_i)\sim
k_i^{-\gamma}$, at the same time that a different functional
dependence is faced in the case of the out-degree distribution
$P(k_o)$, which can present clear exponential cut-offs. In biology,
transcriptional regulatory networks are characterized by the
reflexive situation, in which an incoming degree distribution that
decays faster than a power law can be observed along with a
scale-free outgoing degree distribution~\cite{Vazquez:2004}. It is
then particularly interesting to see what happens for power law in-
or out-degree distributions when combined to exponential out- or
in-degree ones. In this example, the in-degree distribution is taken
to follow a scale-free form of the type
\begin{equation}
P_i(k)= \left\{
\begin{array}{lr}
P_0 & k=0\\[0.5cm]
\displaystyle{\frac{(1-P_0)}{\zeta(\gamma) k^{\gamma}}} & k \ge 1
\end{array}
\right., \label{eq:33}
\end{equation}
where $\zeta(\gamma)$ is the Zeta Riemann function. The out-degree
distribution is given again by Eq.~(\ref{expPk}). The set of
Eqs.~(\ref{gs00}) is solved numerically and plugged into
Eqs.~(\ref{sizes00}) to get the relative sizes of the edge
components and the results are compared to direct measures of the
edge percolation map on a set of synthetically generated networks
with $N=5\cdot10^5$ nodes. We take $\gamma=2.2$, $P_0^{o}=0.4$ and
vary the average degree $\langle k_i \rangle=\langle k_o \rangle$
until the maximum possible value is reached by adjusting $P_0^{i}$.
This upper boundary in the average degree is due to the fact that,
since $P_0^{i}=1-\langle k_i \rangle\zeta(\gamma)/\zeta(\gamma-1)$,
values above the threshold impose a negative $P_0^{i}$ and are not
realizable. The theoretical value for this threshold is
$\zeta(\gamma-1)/\zeta(\gamma)$.

Once again, our predictions compare extremely well with the measures
on the simulated networks, see Fig.~\ref{fig:2}. Interestingly, and
in contrast to what was obtained in the previous example, the edge
percolation map changes dramatically depending on the average
degree. For small values --but big enough to ensure that the system
is in the percolated phase, $\langle k_i \rangle
>1$--, the edge in-component, ICE, and the in-interface, ITF, are
predominant. However, the rest of edge components grow steadily with
the average degree while those reach a maximum and then decay to
eventually disappear at the average degree threshold, so that for
high values of the average degree the edge strongly connected
component, SCE, and the out-interface, OTF, dominate.

\section{Interfaces}
\label{Sec:int}

Interfaces arise as distinctive elements of the edge percolation
map. From the analytical computations one sees that the interfaces
are also giant components. Furthermore, their size could be much
larger than that of the ICE and OCE, for instance as shown in
Figs.~\ref{fig:1} and \ref{fig:2}. In the particular case of the
completely uncorrelated networks with exponential in- and out-degree
distributions given by Eq.(\ref{expPk}), the relative sizes of the
interfaces as compared to that of the pure components can be
calculated analytically and found to be
\begin{equation}
\frac{g_{itf}}{g_{ice}}=\frac{g_{otf}}{g_{oce}}={\langle k_i
\rangle}^2-1. \label{square}
\end{equation}
\begin{figure}[t]
\epsfig{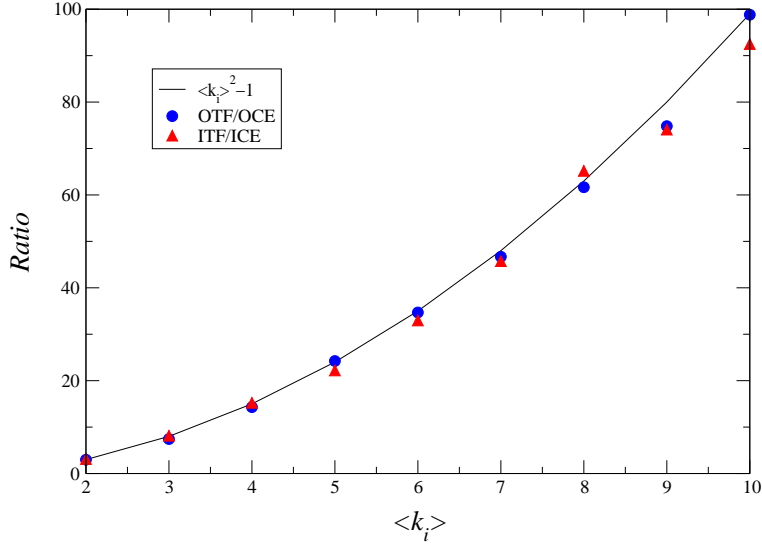}
  \caption{(color online). Ratio of the relative sizes of the giant interfaces to the edge in- and out-components as as a
  function of the average degree. Simulation results (dots)
  correspond to 1-realization networks of size $N=10^5$ vertices with $P_0^{i}=0.4$ and $P_0^{o}=0.8$.
  Solid line correspond to the analytical ratio Eq.~(\ref{square}).}
  \label{fig:3}
\end{figure}
The same relation is numerically seen to happen for the ratio
between the out-interface and the edge out-component of the second
case study where the in-degree distribution was scale-free and the
out-degree distribution exponential. So, for exponential
distributions the ratio of the relative sizes of the giant
interfaces to the corresponding in- or out-component grows
quadratically with the average degree. This result is very
interesting because it suggests that, at least in this case, the
traditional GIN and GOUT components of the node percolation map show
a shallow architecture mainly formed by leaf edges emanating from or
pointing to the SCC. As a consequence, and to give a mental image,
the bow-tie structure of those networks rather becomes a ``hairy
ball''.

All this points to a rich second order fine structure that could
play a central role in the investigation of how topology is related
to functionality. In particular, and apart from the information
contained in both the node and edge percolation maps, the internal
structure of the interfaces, and more specifically the distinction
of leaf edges from connectors, is fundamental in order to assess the
efficiency of the global flow or the risks of bottleneck effects.
Further discussion about the internal structure of interfaces and
possible implications will be provided in a forthcoming work.

\subsection{Internal average degrees}
Interfaces have a hybrid nature from the point of view of node
components. In order to calculate internal average degrees, it is
not clear whether they should be assigned to one node component or
another. If one considers for instance the subset of nodes in the
SCC with all their connections, internal or not, it is found that
\begin{equation}
\frac{\sum_{SCC} k_i}{E}=g_{sce}+g_{itf} \mbox{ , }\frac{\sum_{SCC}
k_o}{E}=g_{sce}+g_{otf},
\end{equation}
where $E$ is the total number of edges in the network. As a
consequence, the detailed balance condition Eq.~(\ref{detailed_dir})
will not be accomplished in general, $\sum_{SCC} k_i\neq\sum_{SCC}
k_o$, except when both interfaces are of equal sizes. The same
happens for the subsets of nodes in the GIN and the GOUT, where from
the point of view of detailed balance there is an excess out- and
in-degree respectively. The interfaces are precisely the responsible
for these imbalances.

We explore once more as a null model that can be fully calculated
analytically the completely uncorrelated network, with no local or
degree-degree correlations. In this situation, the sizes of the main
components in the node percolation map can be expressed as (see
Refs.~\cite{Newman:2002a,Dorogovtsev:2001,Boguna:2005})
\begin{eqnarray}
g_{scc}&=&1-\hat{g}_o(1)-\hat{g}_i(1)+\hat{g}_o(1)\hat{g}_i(1)\nonumber\\
g_{in}&=&1-\hat{g}_o(1)-g_{scc}\nonumber\\
g_{out}&=&1-\hat{g}_i(1)-g_{scc},
\label{sizesnodecomp}
\end{eqnarray}
where $\hat{g}_i(1)$ and $\hat{g}_o(1)$ are the solutions of
Eq.~(\ref{gs00}), like for the edge percolation map. Comparing
Eq.~(\ref{sizesnodecomp}) and Eq.~(\ref{sizes00}), it is found that
the relative sizes in number of nodes of the GIN, GOUT, and SCC are
the same as the relative sizes in number of edges of the ICE+ITF,
OCE+OTF, and SCE respectively. In other words, the average degree of
the whole network is preserved in the different components if the
in- and out-interfaces are assigned to the in- and out-components
respectively. This is in particular valid for the previous examples
of uncorrelated networks with exponential or scale-free in-degree
distribution and exponential out-degree distribution.

\section{Conclusions}
\label{Sec:conc}

We focus on edges instead of nodes to investigate analytically how
they organize in the percolated phase of purely directed random
networks. Interfaces of edges are found to bridge the main
components of the node percolation map. The general case of local
and degree-degree correlations is formally solved and the relative
sizes of the five main giant edge components are characterized
quantitatively. The results for uncorrelated networks are found to
be in very good agreement with direct measures on synthetic
networks, that could present very different edge percolation maps
depending on the in- and out-degree distributions and the average
degree.

The node percolation map is in this way complemented by the edge
percolation map, forming a percolation landscape that gives a more
detailed topological description of the structure of globally
connected systems. The work should not stop here, since results in
this paper seem to point out to the importance of the internal
organization of the interfaces with latent implications at the level
of functional properties. So, the analysis presented in this work
uncovers a new aspect potentially relevant not only for the
structure of directed networks but most importantly for their
functionality. Generally, the SCC processes input into output so
that interfaces become unavoidable bridges that could determine the
effectiveness or the robustness of functional performance.

In this work, we have restricted to purely directed networks, a good
approximation in many cases where flow or transport, when present in
both directions, is asymmetric. Nevertheless, the same ideas can be
extended to semi-directed networks, the most general and realistic
ones. For those, analytical calculations could be a bit more
intricate due to the non-trivial correlations associated to
reciprocity.

\begin{acknowledgments}
We thank Mari{\'a}n Bogu{\~n}{\'a} for helpful discussions. This
work has been financially supported by DELIS under contract FET Open
001907 and the SER-Bern under contract 02.0234.
\end{acknowledgments}

%\bibliographystyle{apsrev}
%\bibliography{ref}

\end{document}